\documentclass[journal]{IEEEtran}
\usepackage{cuted} 
\usepackage{comment}

\title{Simphony: An open-source photonic integrated circuit simulation framework}

\usepackage{cite}
\usepackage{graphicx}
\usepackage{float}
\usepackage{caption}
\usepackage{listings}
\usepackage{xcolor}
\usepackage{amsmath}
\usepackage{mathdots}
\usepackage{authblk}
\usepackage{soul}
\usepackage{xcolor}
\usepackage{ulem}
\definecolor{codegreen}{rgb}{0,0.6,0}
\definecolor{codegray}{rgb}{0.5,0.5,0.5}
\definecolor{codepurple}{rgb}{0.58,0,0.82}
\definecolor{backcolour}{rgb}{0.95,0.95,0.92}
 
\lstdefinestyle{pystyle}{
    backgroundcolor=\color{backcolour},   
    commentstyle=\color{codegreen},
    keywordstyle=\color{magenta},
    numberstyle=\tiny\color{codegray},
    stringstyle=\color{codepurple},
    basicstyle=\ttfamily\footnotesize,
    breakatwhitespace=false,         
    breaklines=true,                 
    captionpos=b,                    
    keepspaces=true,                  
    numbersep=5pt,                  
    showspaces=false,                
    showstringspaces=false,
    showtabs=false,                  
    tabsize=2
}
\begin{document}

\author[1]{Sequoia Ploeg*\thanks{* These authors contributed equally to this work.}}
\author[1]{Hyrum Gunther*}
\author{Ryan M. Camacho$^{\dag}$\thanks{$^{\dag}$ Address correspondence to camacho@byu.edu}}
\affil[1]{Department of Electrical and Computer Engineering, Brigham Young University}

\maketitle

\begin{abstract}
We present Simphony, a free and open-source software toolbox for abstracting and simulating photonic integrated circuits, implemented in Python. The toolbox is both fast and easily extensible; plugins can be written to provide compatibility with existing layout tools, and device libraries can be easily created without a deep knowledge of programming.  We include several examples of photonic circuit simulations with novel features and demonstrate a speedup of more than 20x over a leading commercially available software tool.
\end{abstract}


\section{Introduction}

\IEEEPARstart{S}{ilicon} photonics is a rapidly growing industry that uses electronic integrated circuit fabrication technologies to produce industry-grade photonic integrated circuits (PICs) at low cost and high volume 
\cite{bogaerts_silicon_2018, bogaerts_design_2014}. 
Silicon photonic technologies have been largely driven by the communications industry, but also find applications in sensing, computing, and quantum information processing by enabling high data transmission rates and controlled manipulation of light waves 
\cite{bogaerts_silicon_2018}. 

As the silicon photonics industry grows and the demand for PICs increases, it is increasingly important for designers to have access to software design tools that can accurately model and simulate PICs in a first-time-right framework. Simulating PICs is a resource- and time-intensive process. Owing to the long wavelengths of photons relative to electrons, photonic device simulation requires solving Maxwell's equations with far less abstraction than electronic circuit components. Once devices have been simulated and bench-marked, however, compact models representing the phase and amplitude response functions of individual components may be stitched together to form functioning circuits. Several commercial tools exist to perform these functions, such as Aspic, the formerly open-source IPKISS (now part of Luceda's Caphe), and Lumerical's INTERCONNECT \cite{photontorch}. However, they are often expensive (on the order of thousands of dollars per year) and limited in the variety and type of photonic devices than can be simulated. Furthermore, there is often a lack of standardization among platforms that in many cases prevents interoperability between tools\cite{bogaerts_design_2014}.

In this paper we present a free and open-source, software-implemented simulation tool for linear PICs (documentation and downloads are available at  github.com/BYUCamachoLab/simphony or pip installable via the Python Package Index). Our toolbox, which we name Simphony, provides fast simulations for PICs and allows for the integration of device compact models that may be sourced from a variety of platforms. While providing an easy-to-use and consistent syntax for describing PICs through code, Simphony also provides a framework that allows end users to easily wrap or add their own custom components. This extensibility is achieved by cascading device scattering parameters, or S-parameters, for each component using sub-network growth algorithms, a known and common practice in microwave/radio-frequency (RF) engineering as well as in other PIC simulation software \cite{Leijtens_S_matrix, photontorch}. Benchmark testing of Simphony against Lumerical INTERCONNECT indicates a speedup of approximately 20x. 

Simphony is designed to give more people access to free and open-source software and we foresee it being a very useful tool for a variety of researchers and educators, especially those lacking access to commercial tools.

Simphony also includes optional interoperability with the SiEPIC-Tools extension for KLayout, an open-source, general-purpose layout software \cite{chrostowski_design_2016}.  To accomplish this, Simphony has the ability to parse circuit descriptions created by SiEPIC-Tools thereby allowing simulations to be run directly on circuits designed in KLayout.

The remainder of this paper is organized as follows: In Section 2 we present Simphony's design flow and working principles, including a brief review of the principles and mathematics behind scattering parameter matrices and the standard way of representing compact models.  Section 3 provides several examples of simulations of PICs by Simphony, illustrating its novel functionality and comparing both the accuracy and speed of the simulations to Lumerical's INTERCONNECT, one example of commercial software. Section 4 concludes with a brief discussion and future plans for Simphony.

\section{PIC Design Flow and Working Principles}
Photonic circuits are often much more difficult to simulate than electronic circuits owing to the complex scattering of light in waveguides. Often the only way to accurately simulate photonic circuits is to use computationally expensive methods such as Finite-Difference Time Domain (FDTD) or Finite Element Methods (FEM),  which numerically solve Maxwell’s equations in discretized regions across space and time \cite{oskooi_meep}. Though FDTD and similar methods produce accurate results, simulations can be computationally prohibitive when scaling the size and complexity of the circuit.  As many modern silicon photonic circuits contain 10's to 100's of devices, there is a need for novel simulation techniques.  

\begin{figure}[t]
\centering
\includegraphics[width=0.48\textwidth]{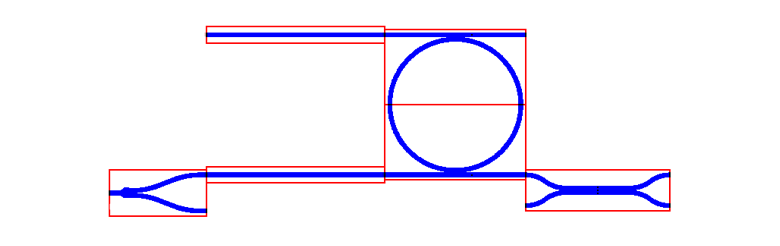}
\caption{The block driven-design methodology used by many commercial programs that is also adopted by Simphony. For example, this photonic circuit can be divided into a few once-simulated, reusable components.}
\label{fig:block_methodology}
\end{figure}

To address this problem, many designers have moved to a block-driven design methodology \cite{chrostowski_design_2016}, as shown in Figure \ref{fig:block_methodology}. Rather than simulating large, complex circuits, they run simulations of basic photonic components such as waveguides, splitters, and couplers, and then combine the outputs of those simulations in ways that accurately predict the behavior of the whole circuit. Simulations of individual components take far less time than the simulation of a complex circuit and the results of small-scale component simulations can be saved and used more than once. Designers that follow a block-driven methodology typically layout and simulate many basic components, storing the results of those simulations in a compact model library (CML). Recently, researchers have also used block-based methods to represent components and circuits as nodes and connections within a neural network, as in Photontorch \cite{photontorch}.  The primary function of Simphony is to produce accurate circuit simulations given compact models and a description of their interconnections.

\subsection{Implementation of Scattering Parameters}
\begin{figure}[t]
\centering
\includegraphics[width=0.48\textwidth]{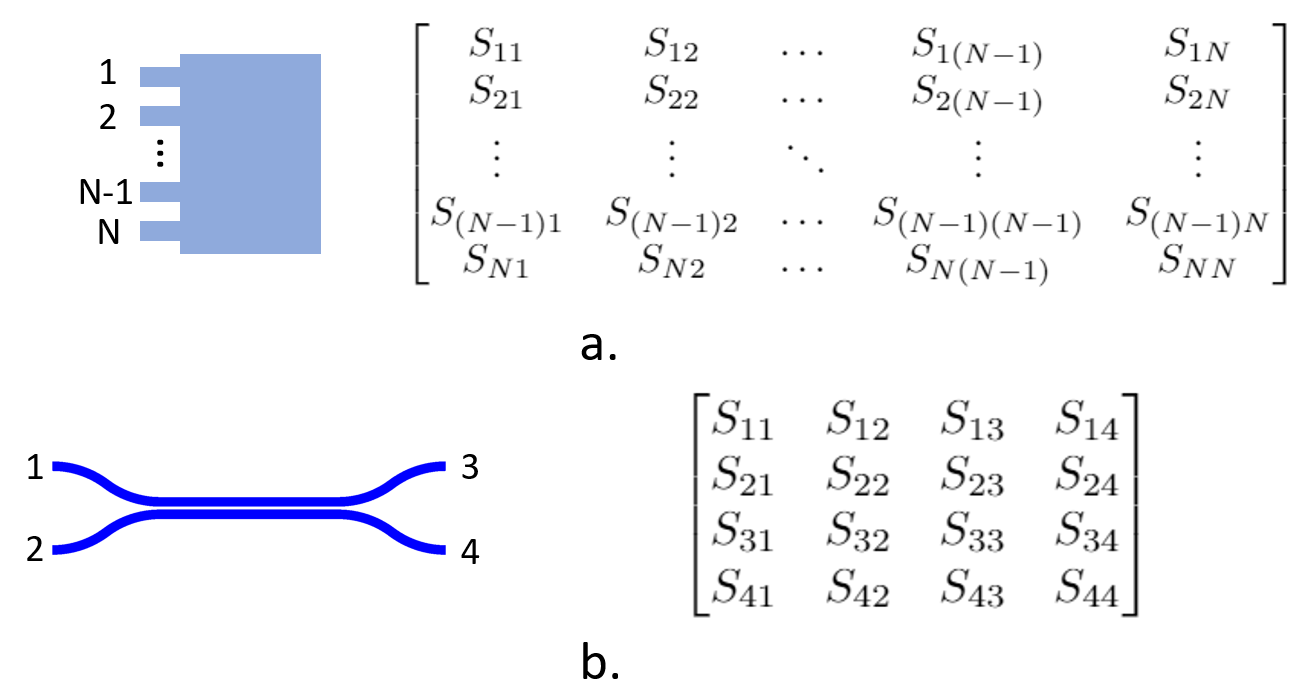}
\caption{Illustration of a scattering matrix for a photonic device. a) A generic $N$-port device can be represented by an $N\!\times \!N$ scattering matrix. A non-diagonal element, $S_{ij}$, of the matrix describes the transmission of the electric field from input $j$ to output $i$. A diagonal element, $S_{ii}$, describes the back reflection of the E-field for port $i$. b) A common photonic component is the directional coupler, which couples light by bringing two waveguides close together and then separating again. As it is a 4-port device, it can be represented with a $4\!\times \!4$ scattering matrix.}
\label{fig:s_matrix_illustration}
\end{figure}

A compact model's scattering parameters describe the phase and amplitude relationship between ports of an abstracted device or circuit, as illustrated in Figure \ref{fig:s_matrix_illustration}. Well-known to RF and microwave engineers, they allow a designer to model system behavior without dealing with each component's inner workings. 
 
In Simphony, a device or component with $N$ connection points, called ports or pins interchangeably, is represented by an $N\!\times \!N$ scattering parameter matrix $S$. 
An element $S_{ij}$ of the matrix represents the ratio of the field amplitude (not power) entering port $j$ and exiting port $i$.  The diagonal elements of the scattering matrix $S_{ii}$ are the ratios of the field entering port $i$ and exiting back out the same port, representing the back reflection of the device for that specific port.

Simphony is also designed to handle components with frequency-dependent S-parameters, allowing for frequency sweep simulations. This also allows for time-dependent simulations, that, while not yet implemented in Simphony, can be performed using impulse response functions.  For many of the compact models provided in Simphony's default library (presimulated components with fixed parameters), compact models may contain S-parameters for  $N_f$ frequency points, making the scattering matrix of size $N_f\!\times \!N\!\times \!N$. When simulations over a range of frequencies are run, S-parameters can be easily interpolated for each compact model given that they are slowly varying (for example, there are no sharp resonances).

\subsection{Sub-Network Growth}
\begin{figure}[t]
\centering
\includegraphics[width=0.48\textwidth]{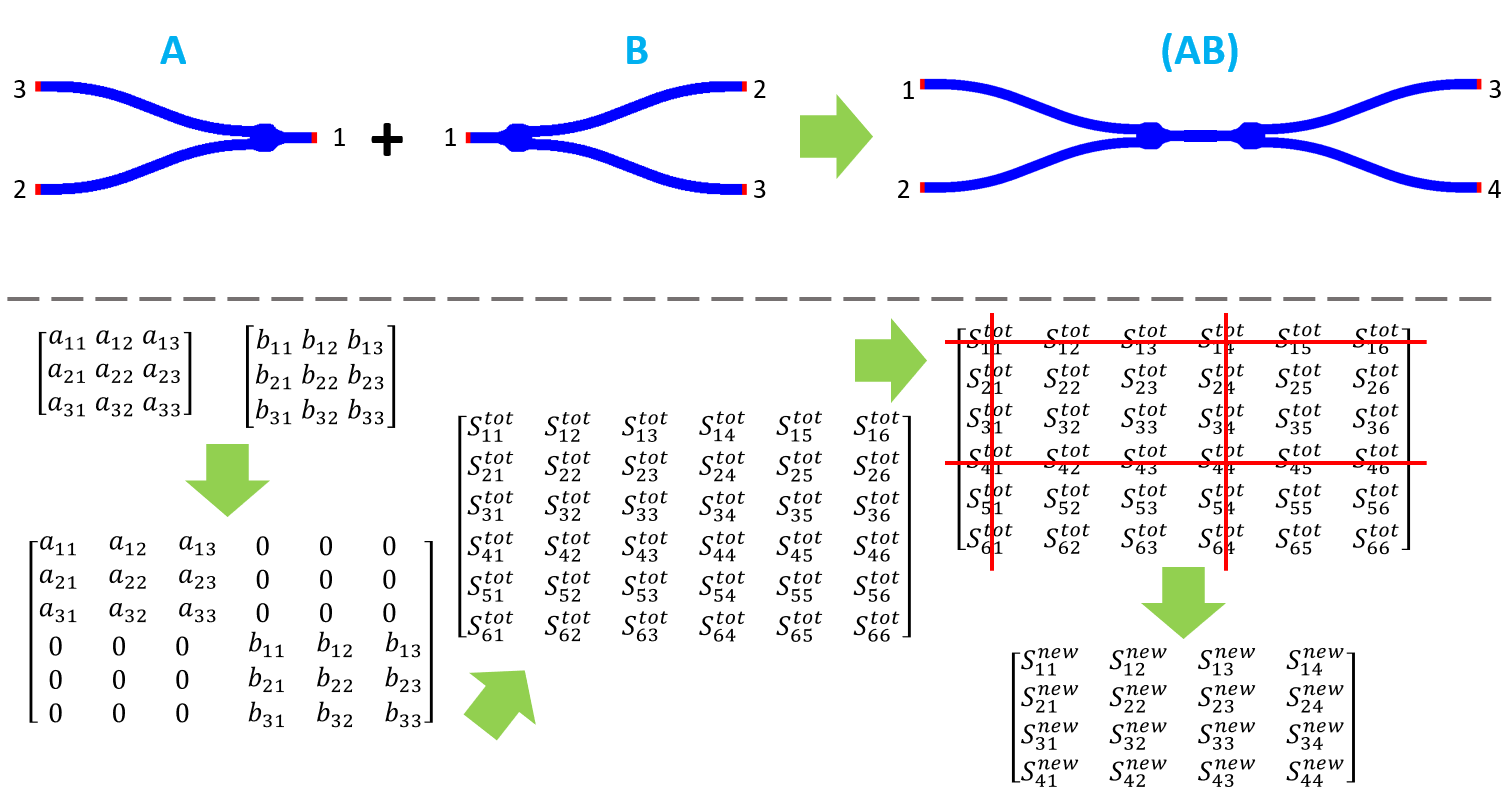}
\caption{An example of sub-network growth. Two y-splitters, $A$ and $B$, each a 3-port component, are to be connected at port 1 on both components. The result will be the single 4-port device shown in the top right. To connect the components, the scattering matrices of each component are placed as the diagonal of a composite matrix. The interconnection operation described above is performed on each element of the composite matrix, resulting in $\protect\begin{bmatrix}S^{tot}\protect\end{bmatrix}$. Lastly, the rows and columns corresponding to port 1 on each y-splitter are removed, resulting in the $4\!\times \!4$ scattering matrix of the 4-port device.}
\label{fig:subnetwork_growth}
\end{figure}

Simphony combines S-parameters from individual components by performing operations to connect components as dictated by a netlist. A netlist is a list representation of the components and their respective connections within a layout \cite{schematic_driven_lukas}. The combined S-parameters are then used to perform simulations. An individual component's scattering matrix $S$ describes the relationship between the complex field amplitudes at the component's input and output ports, $A^+$ and $A^-$, respectively:
\[
    \begin{bmatrix}
    A_1^{-} \\
    \vdots \\
    A_k^{-} \\
    \vdots \\
    A_l^{-} \\
    \vdots \\ 
    A_N^{-}
    \end{bmatrix} = 
    \begin{bmatrix}
    S_{11} & & & \dots & & & S_{1N} \\ 
     & \ddots & & & & \iddots & \\
     & & S_{kk} & \dots & S_{kl} & & \\
    \vdots & & \vdots & & \vdots & & \vdots \\
     & & S_{lk} & \dots & S_{ll} & & \\
     & \iddots & & & & \ddots & \\
    S_{N1} & & & \dots & & & S_{NN}
    \end{bmatrix} \cdot
    \begin{bmatrix}
    A_1^{+} \\
    \vdots \\
    A_k^{+} \\
    \vdots \\
    A_l^{+} \\
    \vdots \\ 
    A_N^{+}
    \end{bmatrix}
\]
Simphony combines S-parameters using a ``sub-network growth" algorithm \cite{filipsson_new_1981}, shown graphically in Figure \ref{fig:subnetwork_growth}. The sub-network growth routine operates on scattering matrices in two ways: (1) connecting ports within a single scattering matrix (self-connections), or (2) connecting ports among distinct scattering matrices. 

For the first case, suppose an $N$-port component has self-connections at only two of its ports, $k$ and $l$.  We then have the relationships $A_k^- = A_l^+$ and  $A_l^- = A_k^+$, and we may rearrange rows of $S$, to obtain the following relationships:
\clearpage

\begin{strip}
\begin{align}
    A_l^+ = \frac{(1-S_{lk})(1-S_{kl})}{1-S_{kl}-S_{lk}+S_{kl}S_{lk}-S_{kk}S_{ll}}
    \sum_{i=1, i\neq k,l}^N \left(\frac{S_{ki}}{1-S_{kl}} + \frac{S_{kk}S_{li}}{(1-S_{kl})(1-S_{lk})} \right) A_i^+
\end{align}

\begin{align}
    A_k^+ = \frac{(1-S_{lk})(1-S_{kl})}{1-S_{kl}-S_{lk}+S_{kl}S_{lk}-S_{kk}S_{ll}}
    \sum_{i=1, i\neq k,l}^N \left(\frac{S_{li}}{1-S_{lk}} + \frac{S_{ll}S_{ki}}{(1-S_{kl})(1-S_{lk})}\right) A_i^+
\end{align}
\end{strip}
By connecting ports $k$ and $l$ in a single component, the $N$-port scattering matrix $S$ becomes an ($N-2$)-port matrix, $S^{tot}$.  
The S-parameters $S_{ij}^{tot}$ can then be expressed as:
\begin{strip}
\begin{align}
    S_{ij}^{tot} = S_{ij} + \frac{S_{il}S_{kj}(1-S_{lk})+S_{il}S_{kk}S_{lj}+S_{ik}S_{lj}(1-S_{kl})+S_{ik}S_{ll}S_{kj}}{1-S_{kl}-S_{lk}+S_{kl}S_{lk}-S_{kk}S_{ll}}.
\end{align}
\end{strip}

In the second case, the S-parameters of two separate components, $A$ and $B$, with scattering matrices $S_A$ and $S_B$, are combined using one port from each component. This is accomplished by creating a new matrix, $S$, from the two scattering matrices using the following rule:
\begin{align}
    S=
    \begin{bmatrix}
    S_A & 0 \\
    0 & S_B
    \end{bmatrix}
\end{align}
and then performing the above interconnection calculation for $S^{tot}$ on each element of $S$ for the two ports that are to be connected. Simphony implements sub-network growth using several functions within scikit-rf, an open-source Python package for RF and microwave applications.  Simphony's simulator traverses the generated netlist and operates on the individual scattering matrices and port numbers according to Eqs 1-3. The result is a single scattering matrix with non-diagonal elements representing complex transmission amplitudes from an input port to an output port of the circuit and diagonal elements representing reflection of light into and out of a given port. 

\subsection{Included Models and S-Parameter Generation}
Simphony includes a basic set of common compact models based on the SiEPIC EBeam Process Design Kit (PDK) \cite{schematic_driven_lukas}. These models include grating couplers, waveguides, y-branches, half-rings of specific radii, directional couplers, and waveguide terminators, all operating within approximately the 1500-1600 nanometer range. Simphony also includes a library of components based on linear regression models, discussed later. These models include waveguides, directional couplers, and half-rings of arbitrary widths, thicknesses,  coupling gaps, and radii, also around the 1450-1650 nanometer range. These models are among the most useful contributions to the simulation toolbox, as they allow for rapid simulation without the need for new and computationally expensive full-field simulations. In cases where a desired component is not available within Simphony's libraries, users can turn to an alternative free and open-source software that performs FDTD simulations to generate S-parameters for use within Simphony, such as MEEP \cite{oskooi_meep}. It is our hope that over time, the open-source community will also continue to contribute models to Simphony and further extend the software and its simulation capabilities.

\section{Simulation Examples}
In this section we present simulation results produced by the toolbox to demonstrate its features and speed. We compare the results to those produced by commercial products (Lumerical INTERCONNECT) and also compare the speed of the simulations from both toolboxes.

\subsection{Mach-Zehnder Interferometers}
\begin{figure}[ht]
\centering
\includegraphics[width=0.4\textwidth]{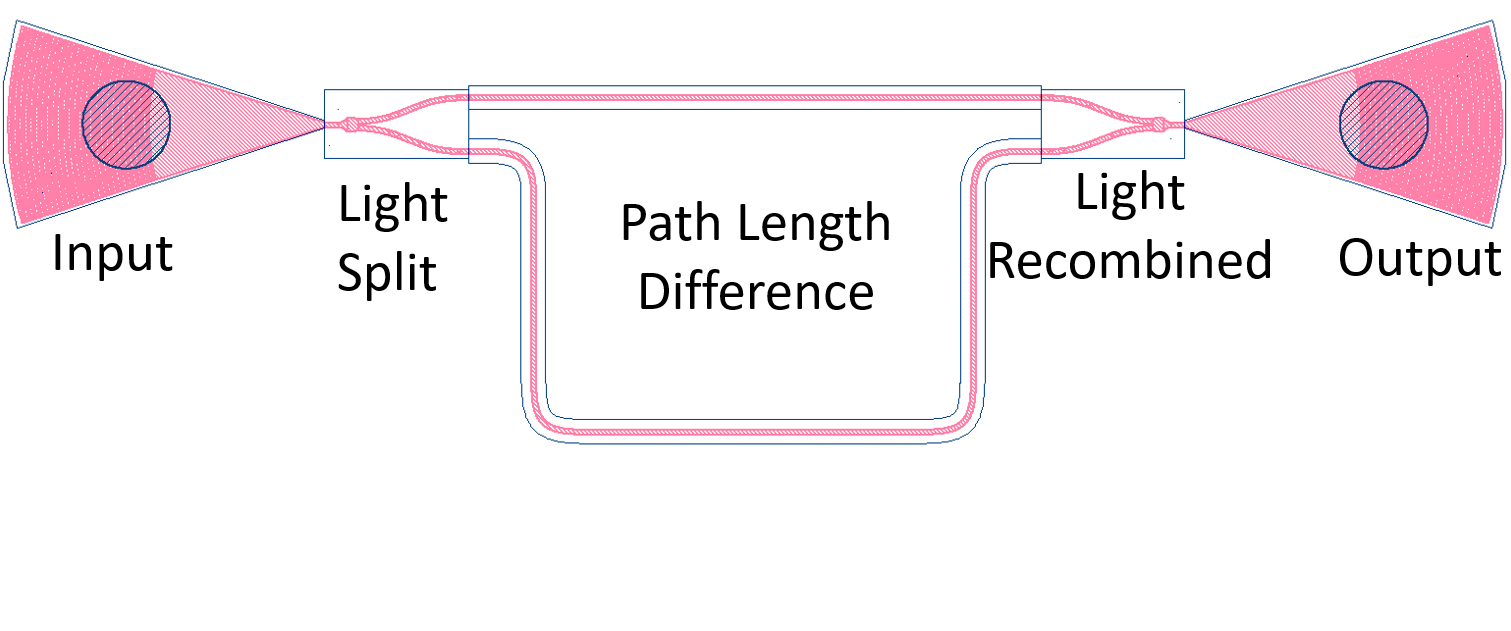}
\caption{A labeled screenshot of a Mach-Zehnder interferometer as designed in KLayout using the SiEPIC EBeam PDK. Light is coupled into the device via a grating coupler and then split by a 50/50 splitter. The split light travels down two paths of different lengths before being recombined by another 50/50 splitter and output through a second grating coupler. The difference in path length causes the recombining light to interfere as the input wavelength is varied.}
\label{fig:MZI_diagram}
\end{figure}

As a first example, we demonstrate modeling an integrated photonics Mach-Zehnder interferometer (MZI) (see Figure \ref{fig:MZI_diagram}). This circuit was first designed fully in KLayout with SiEPIC-Tools and using compact models from the SiEPIC EBeam PDK. Using the SiEPIC-Tools package allows for KLayout to act as a drag-and-drop interface for connecting components, thus enabling the circuit to be designed with a layout-driven methodology \cite{chrostowski_design_2016}.  To simulate this circuit, Simphony simply parses the same netlist that SiEPIC-Tools already generates for use in Lumerical INTERCONNECT simulations. This addition to the SiEPIC toolbox allows Simphony to be integrated natively within KLayout and simulations run on layouts created there. Since the models found in this circuit are also included in the default distribution of Simphony, this compatibility with the SiEPIC toolbox allows Simphony to be easily incorporated into design and simulation workflows for layouts created in KLayout.

Not only does Simphony easily extend existing design tools but simulations can also be scripted directly using Python without layout tools. To demonstrate this, we present a code walkthrough where we construct the same MZI circuit by importing Simphony and declaring the necessary components and their connections. We use two grating couplers, two y-branch couplers, and two waveguides of differing lengths, as shown in Figure \ref{fig:MZI_diagram}. Compact models are provided as instantiable objects within Simphony's ``library" submodule. To define each waveguide's length, a required parameter for calculating S-parameters, we simply include it as an argument upon instantiation (note that the online documentation for all models included in Simphony's default libraries lists the available parameters for each model along with pertinent information, such as units). In Listing 1 we instantiate models for grating couplers, y-branches, a 50-micron waveguide, and a 150-micron waveguide in preparation for including them within a circuit.

\begin{lstlisting}[language=Python, caption=Initializing the components needed to construct a simple MZI]
# Declare the models used in the circuit
from simphony.library import siepic
grating = siepic.ebeam_gc_te1550()
y = siepic.ebeam_y_1550()
wg150 = siepic.ebeam_wg_integral_1550(length=150e-6)
wg50 = siepic.ebeam_wg_integral_1550(length=50e-6)
\end{lstlisting}

Once the models to be used have been instantiated with the appropriate parameters, we can add them to a circuit. As models are added to the circuit, Simphony creates unique component instances behind the scenes that reference a given model only to calculate S-parameters; hence, a model that occurs multiple times in a circuit need only be instantiated once but can be added multiple times to a circuit, representing multiple occurrences of that model within a circuit. Since components are unique within a circuit, Simphony allows each added component to be named with a string identifier for easier referencing later on (for example, when defining connections between ports of adjacent devices). Also note in Listing 2 that models are added into a Subcircuit object; subcircuits may be simulated on their own or embedded within other subcircuits, allowing for even higher levels of abstraction or the scripting of custom, parameterized subcircuits.

\begin{lstlisting}[language=Python, caption=Instantiating devices based on models within the circuit.]
# Create the circuit, add all individual instances
from simphony.netlist import Subcircuit
circuit = Subcircuit('MZI')
e = circuit.add([(grating, 'input'),
                 (grating, 'output'),
                 (y, 'splitter'),
                 (y, 'recombiner'),
                 (wg150, 'wg_long'),
                 (wg50, 'wg_short')])
\end{lstlisting}

All models have default port names stored within the \texttt{pins} class attribute (for included libraries, these can found in Simphony's online documentation). Once a model has been added to a circuit and become a component, however, its ports can be renamed independent of its parent model for ease of referencing later on, as shown in Listing 3. The purpose of this feature is to allow scripted circuits to be more human-readable and therefore easier to debug, especially as the complexity and number of components within a circuit increases.

\begin{lstlisting}[language=Python, caption={Renaming ports, called ``pins" in Simphony, of devices within the circuit for easier access later.}]
# Individual pins can be renamed:
circuit.elements['input'].pins['n2'] = 'input'
circuit.elements['output'].pins['n2'] = 'output'
# Or, all pins can be renamed simultaneously, by default order:
circuit.elements['splitter'].pins = ('in1', 'out1', 'out2')
circuit.elements['recombiner'].pins = ('out1', 'in2', 'in1')
\end{lstlisting}

Next, we define the connections. This is done in Simphony by passing a list of tuples, each tuple listing (in order) a component, the port to make the connection from, the connecting component, and the corresponding port to make the connection to, as shown in Listing 4.

\begin{lstlisting}[language=Python, caption=Defining the connections of an MZI using a list of tuples.]
circuit.connect_many([
    ('input', 'n1', 'splitter', 'in1'),
    ('splitter', 'out1', 'wg_long', 'n1'),
    ('splitter', 'out2', 'wg_short', 'n1'),
    ('recombiner', 'in1', 'wg_long', 'n2'),
    ('recombiner', 'in2', 'wg_short', 'n2'),
    ('output', 'n1', 'recombiner', 'out1'),
])
\end{lstlisting}

Finally, to run the simulation, the desired simulation type is imported and the circuit passed as a parameter. Other parameters specific to certain simulations, such as frequency range, units, and the number of intermediary points to be calculated, can be found in the documentation.

\begin{lstlisting}[language=Python, caption=Simulating our simple MZI.]
from simphony.simulation import SweepSimulation
simulation = SweepSimulation(circuit, 1500e-9, 1600e-9)
result = simulation.simulate()
freq, power = result.data('input', 'output')
\end{lstlisting}

\begin{figure}[t]
\centering
\includegraphics[width=0.4\textwidth]{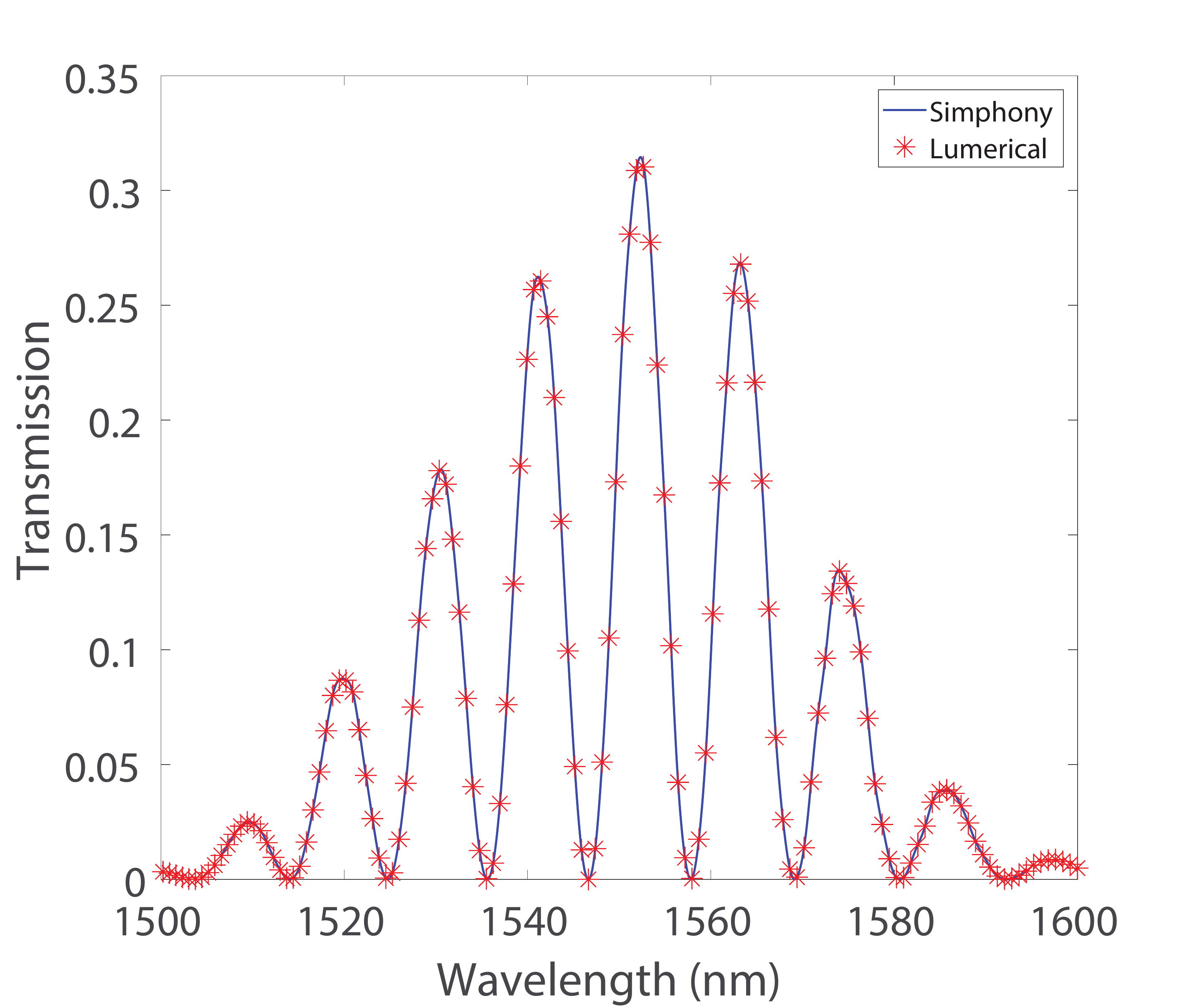}
\caption{The results of a Simphony simulation and a Lumerical INTERCONNECT simulation of the MZI shown in Figure \ref{fig:MZI_diagram}. }
\label{fig:MZI_data}
\end{figure}

\begin{figure}[b]
\centering
\includegraphics[width=0.48\textwidth]{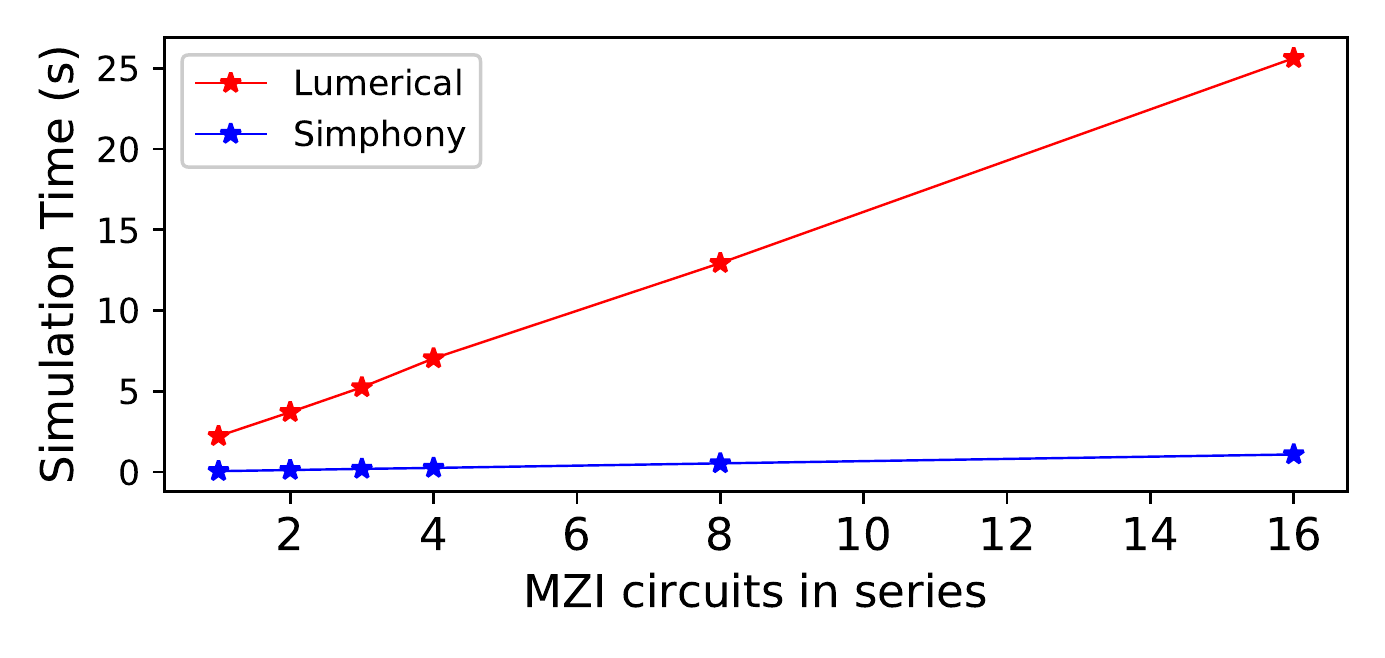}
\caption{A simulation time comparison between Simphony and Lumerical INTERCONNECT for MZI's cascaded in series, with test cases (starred values) connected by lines. As the number of components in the circuit increases linearly, the processing time required for simulation increases proportionally with it. Simphony completes the simulation approximately 20x faster than INTERCONNECT.}
\label{fig:timing_comparison}
\end{figure}

The results of the simulation are shown in Figure \ref{fig:MZI_data}.  Two features are noteworthy. First, we note that the simulation matches the results obtained by simulating the same circuit using the commercial software Lumerical INTERCONNECT, a very useful and accurate simulation tool. We also note, as shown in Figure \ref{fig:timing_comparison}, that the simulated circuit runs approximately 20x faster within the Simphony toolbox than it does in INTERCONNECT, one example of commercial software. This is an especially useful feature when scripting simulations of large circuits with several varying parameters. In benchmarking our speed tests, we used a laptop with an Intel Core i7 2.5GHz quad-core CPU running Ubuntu 18.04. We used the Lumerical 2019b version of INTERCONNECT and Python 3.6.8 for Simphony. INTERCONNECT was invoked through its built-in scripting window on netlists comprised of components already added to INTERCONNECT's library. The circuits were MZIs similar to that shown in Figure \ref{fig:MZI_diagram}, with each successive MZI added in series being connected by an additional waveguide. The simulation times shown are an average over 10 runs per circuit.

\subsection{The Green Machine}

\begin{figure}[t]
\centering
\includegraphics[width=0.48\textwidth]{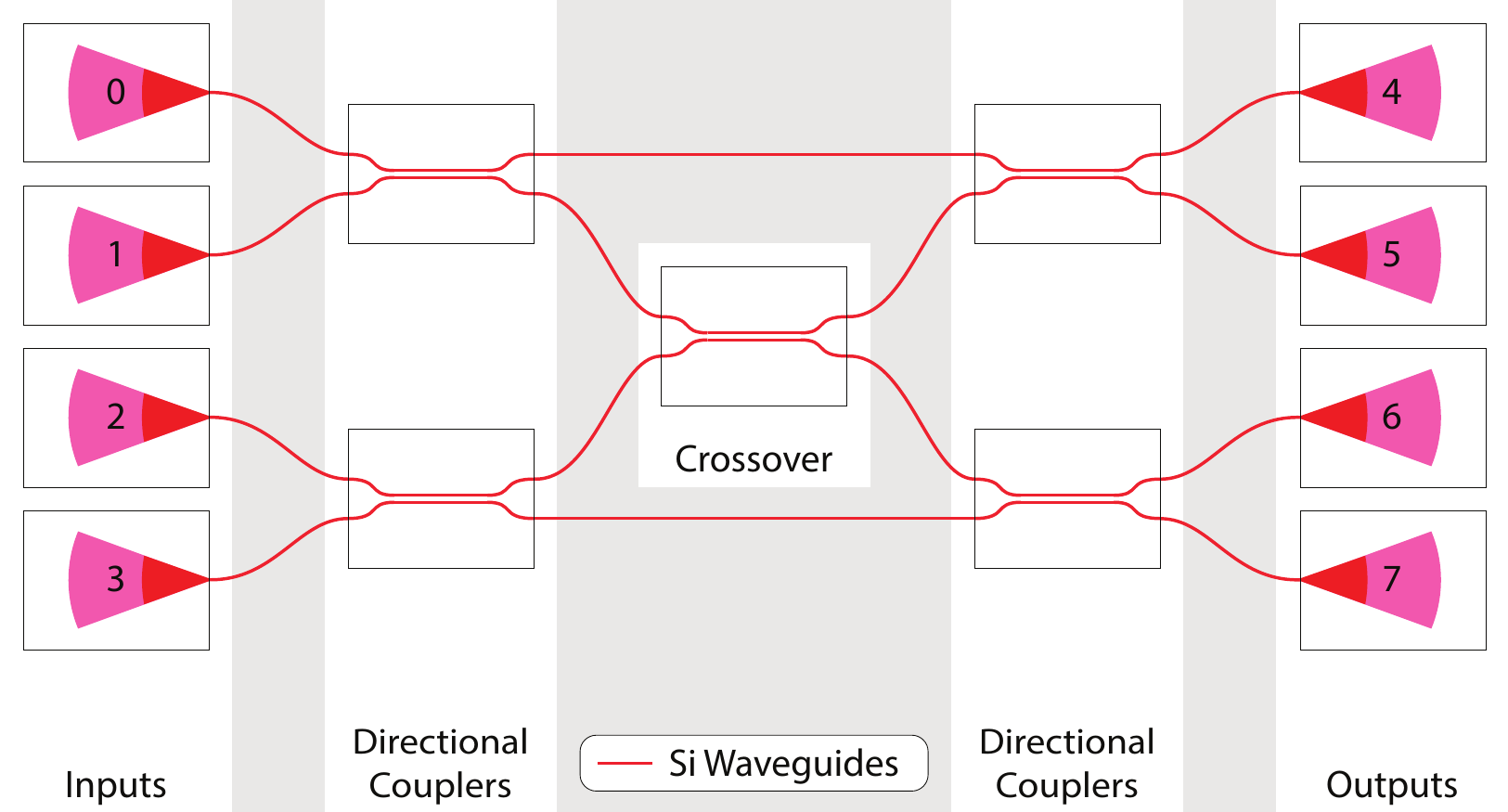}
\caption{A block-style diagram of a PIC device known as a ``Green Machine." This complex structure could take several days to simulate on a supercomputer using FDTD or full-wave analysis techniques. Breaking it into components as shown in this block diagram allows it to be solved in seconds, allowing for faster iterations on the design. The above diagram only contains four basic component models: grating couplers, directional couplers, crossover structures, and waveguides. Device port numbers are labeled on each grating coupler.}
\label{fig:green_machine_circuit}
\end{figure}

As a second simulation example, we simulate a more complicated integrated photonic circuit known in the literature as the ``Green Machine" (GM)\cite{guha_structured_2011}. The GM is a scalable photonic circuit whose topology matches that of the fast Fourier transform (FFT) by mapping the inputs and outputs of various beamsplitter stages in a butterfly fashion.  A 4-port integrated photonic GM circuit is shown in Figure \ref{fig:green_machine_circuit}. Functionally, it performs a Hadamard transform on a binary phase-shift keyed (BPSK) codebook, generating a pulse-position modulation (PPM) based codebook. 

To simulate the Green Machine using Simphony, we divide the photonic circuit into its base components and simulate the entire device using sub-network growth. Importantly, this can be scripted to add an arbitrary number of components using the direct Python interface.  As shown in Figure \ref{fig:green_machine_circuit}, the GM circuit consists of 8 grating couplers, 4 directional couplers, a crossover, and connecting waveguides.

In order to illustrate the flexibility of the Simphony toolbox, the compact models for the grating couplers, waveguides, directional couplers, and the crossover were built using different tools and integrated within Simphony.  The waveguide and grating coupler compact models were included from the SiEPIC EBeam PDK, while the directional couplers and crossover structures were built using a custom simulator based on machine learning techniques\cite{Hammond:19}.

\begin{figure}[t]
\centering
\includegraphics[width=0.48\textwidth]{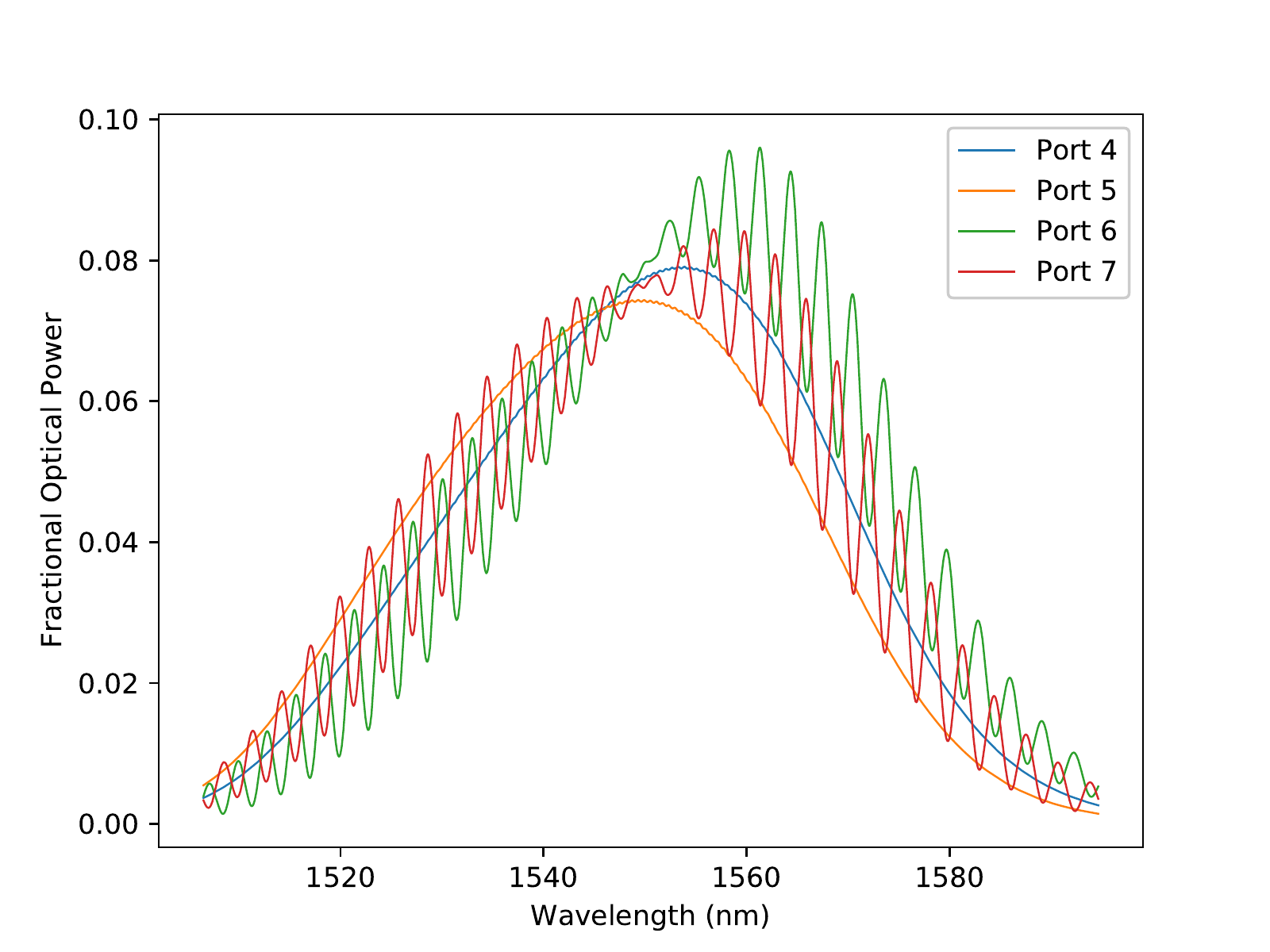}
\caption{Simulation results for the green machine circuit. We used a crossover optimized for a 0/100 crossing ratio at 1550nm. Using compact models of non-ideal components throughout the circuit, the simulation data shows the desired response, or point where the optical power is most evenly split between the four outputs, occurs at that wavelength.}
\label{fig:green_machine_sim}
\end{figure}

Figure \ref{fig:green_machine_sim} shows the result of injecting light into port 1 of the GM circuit and monitoring the outputs of ports 4-7.  Notably, a complicated spectral response function results that might be difficult to unravel without accurate models that describe each of the individual components. As expected, each output port receives a similar amount of power, but it is less obvious how to interpret the complete response. By numerically de-embedding each component using Simphony, however, we can identify the broad spectral envelope resulting from the grating couplers and the oscillations in ports 6 and 7 that result from non-idealities in the crossover.

By analyzing the phase response of the GM circuit at 1550 nm, where the directional couplers and crossovers are designed to have near-ideal 50/50 and 100/0 splitting ratios, we can assess the designed GM's performance.  Table \ref{table:1} shows the ideal phase response of the GM circuit across each input/output combination.

\begin{table}[ht]
\centering
\begin{tabular}{ |c|c|c|c|c| } 
 \hline
  in/out port & 4 & 5 & 6 & 7 \\ 
  \hline
 0 & 0 & $\pi/2$ & $\frac{\pi}{2}$ & $\pi$ \\ 
 \hline
 1 & $\pi/2$ & $\pi$ & 0 & $\pi/2$ \\ 
 \hline
 2 & $\pi/2$ & 0 & $\pi$ & $\pi/2$ \\ 
 \hline
 3 & $\pi$ & $\pi/2$ & $\pi/2$ & 0 \\ 
 \hline
\end{tabular}
\caption{Codewords (relative phases) of the light at each output when light is coupled into some input.}
\label{table:1}
\end{table}

\begin{figure}[tt]
\centering
\includegraphics[width=0.48\textwidth]{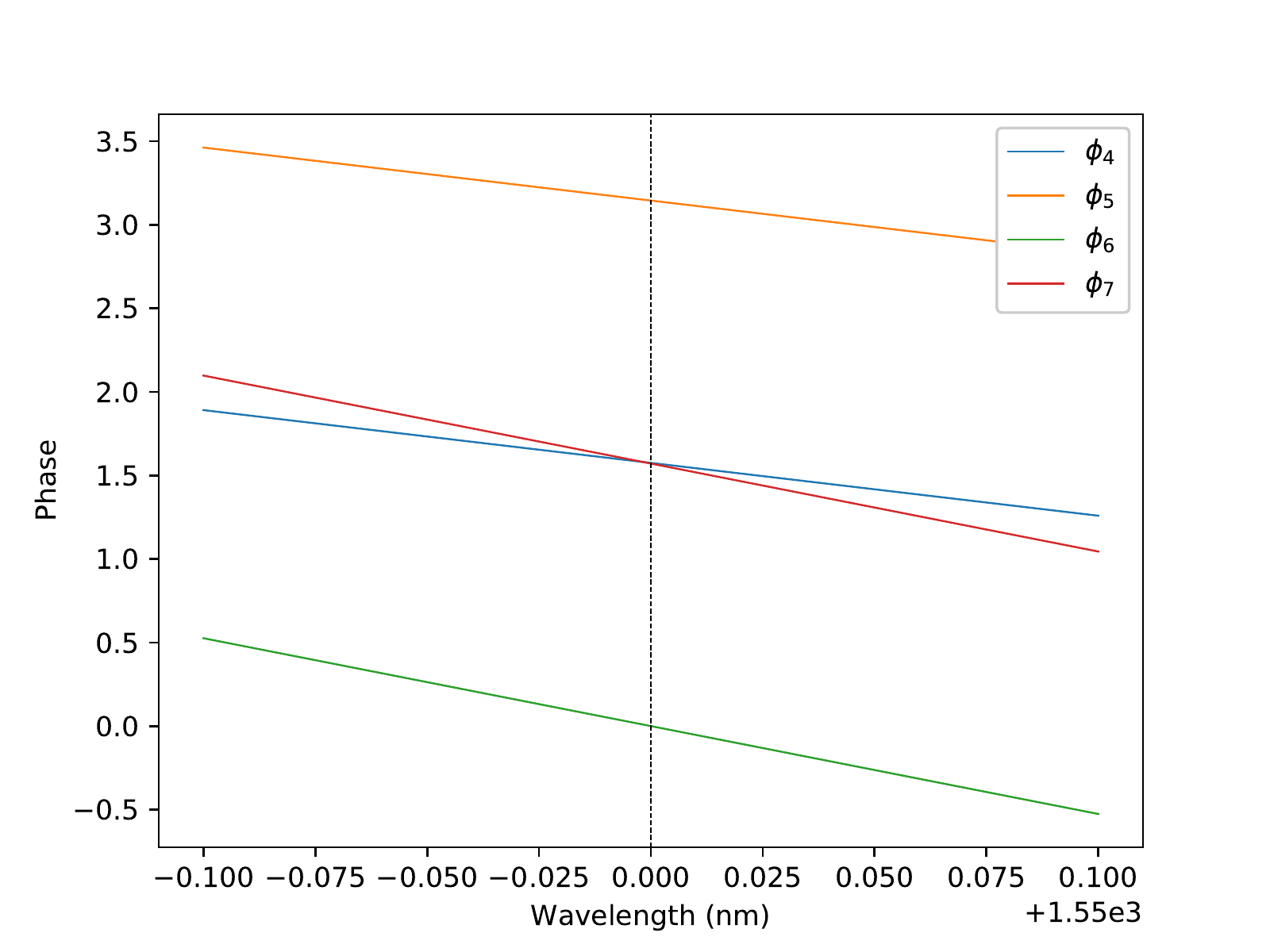}
\caption{GM codewords change as a function of wavelength.}
\label{fig:codewords}
\end{figure}

Figure \ref{fig:codewords} shows the relative phase of the four output ports.  At 1550 nm, the phases almost exactly match the ideal codewords shown in the first column of Table \ref{table:1}.  However, at wavelengths detuned from 1550 nm, the relative phases drift.  This demonstrates the power of Simphony in being able to predict potential errors rapidly during the design process.  

It is also worth noting that in most existing frameworks with presimulated components, parameters (such as waveguide thickness and width) are hard-coded into the compact models. In contrast, as demonstrated in the GM example,  Simphony includes components whose responses are calculated based on linear regression models, thus allowing these parameters, which include waveguide width, thickness, and sidewall angle, to be adjusted and simulation results produced without the computational cost of running an entire analysis for each new set of parameters. This allow us to adjust waveguide lengths, coupler gaps, etc., and get simulation results in seconds rather than days. These models allow us to maintain the block-based circuit description model while also including dynamic, parameterized devices that can be optimized for desired responses. The valid ranges of the parameters for the devices based on linear regression models, designed for use in silicon PICs, can be found in the software documentation.

\subsection{90$^\circ$ Optical Hybrid}
\begin{figure}[ht]
\centering
\includegraphics[width=0.48\textwidth]{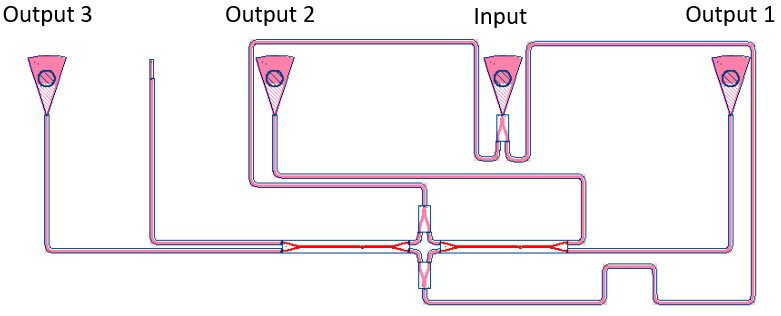}
\caption{A labeled screenshot of a modified 90$^\circ$ Optical Hybrid as designed in KLayout with SiEPIC-Tools. Typically these hybrids will have two inputs and four outputs. In this case we use a single input and split it using a 50/50 splitter, in effect creating two inputs from a single source. These inputs are split and shifted in the directional couplers, creating four quadratures that are all 90$^\circ$ offset from each other. In this particular circuit, three of these quadratures are sent to outputs and one is terminated.}
\label{fig:optical_hybrid}
\end{figure}

This example again illustrates Simphony's operation in conjunction with KLayout, though the capability for scripting it in Python also exists.  In this example, Simphony is used to simulate a 90$^\circ$ optical hybrid. These devices are used in coherent transmission systems to mix an incoming signal with the quadrature states of a local oscillator\cite{guan_compact_2017}. The circuit we simulate here was designed for a specific fabrication run that only allowed for a single input and three outputs per device, so the general design for a 90$^\circ$ optical hybrid was modified to accommodate for this (see Figure \ref{fig:optical_hybrid}). Instead of two separate inputs, the single allowed input was split by a 50/50 y-branch splitter and sent down paths of different lengths to act as the signal and local oscillator. Three of the four quadratures were connected to the available outputs and the fourth was sent to a terminator.

To obtain experimental data to match with the designed circuit, a chip was fabricated at the University of Washington in collaboration with the University of British Colombia and the SiEPIC program on a 150 mm silicon-on-insulator (SOI) wafer with 220 nm thick silicon on 3 $\mu$m thick silicon dioxide and a hydrogen silsesquioxane resist (HSQ, Dow-Corning XP-1541-006). Electron beam lithography was performed using a JEOL JBX-6300FS system operated at 100 keV energy \cite{bojko_electron_2011}, 8 nA beam current, and 500 $\mu$m exposure field size. The silicon was removed from unexposed areas using inductively coupled plasma etching in an Oxford Plasmalab System 100. Cladding oxide was deposited using plasma enhanced chemical vapor deposition (PECVD) in an Oxford Plasmalab System 100.

\begin{figure}[ht]
\centering
\includegraphics[width=0.5\textwidth]{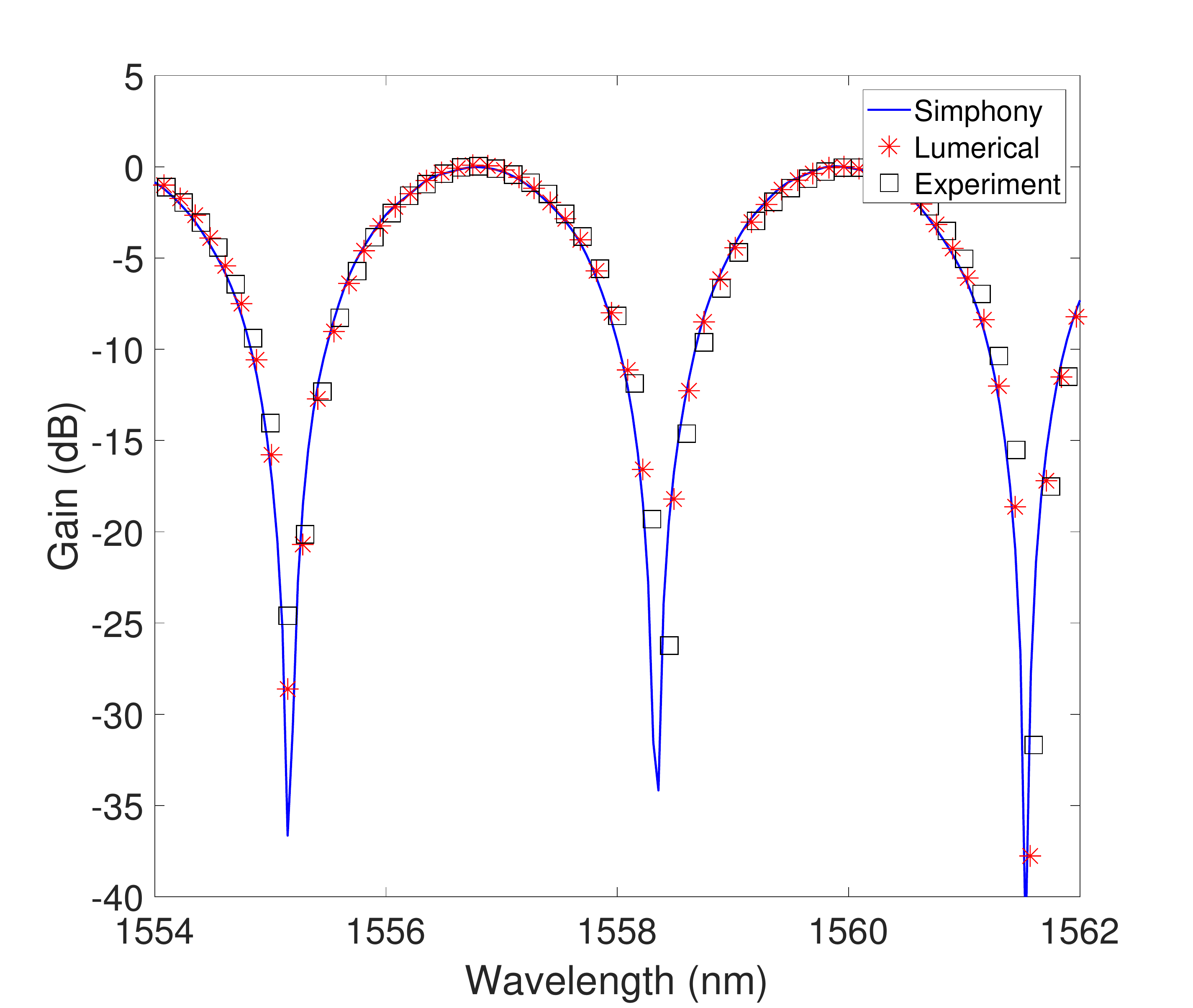}
\caption{Simphony simulation results, Lumerical Interconnect simulation results, and experimental data for Output 1 of the modified optical hybrid circuit in Figure \ref{fig:optical_hybrid}. Simulation data has been shifted laterally to line up with the experimental data.}
\label{fig:optical_hybrid_comparison}
\end{figure}

Figure \ref{fig:optical_hybrid_comparison} shows Simphony simulation results, Lumerical Interconnect simulation results, and measured experimental data from fabricated devices. Both the Simphony and INTERCONNECT simulation data in Figure \ref{fig:optical_hybrid_comparison} has been laterally shifted to line up with the experimental data that is shifted owing to manufacturing variability. Manufacturing variability all but guarantees that separate fabrication runs will result in devices with peaks at differing wavelengths; shifting is acceptable as the feature of interest is the free spectral range (FSR) of the device, or the spacing in wavelength between adjacent optical intensity minima, and not the wavelengths at which maximal interference occurs. The Simphony simulation matches well with both the Lumerical INTERCONNECT simulation results and the experimental data, demonstrating Simphony's usefulness for predicting circuit behavior. 

\subsection{Optical Filtering Circuits}

\begin{figure}[t]
\centering
\includegraphics[width=0.5\textwidth]{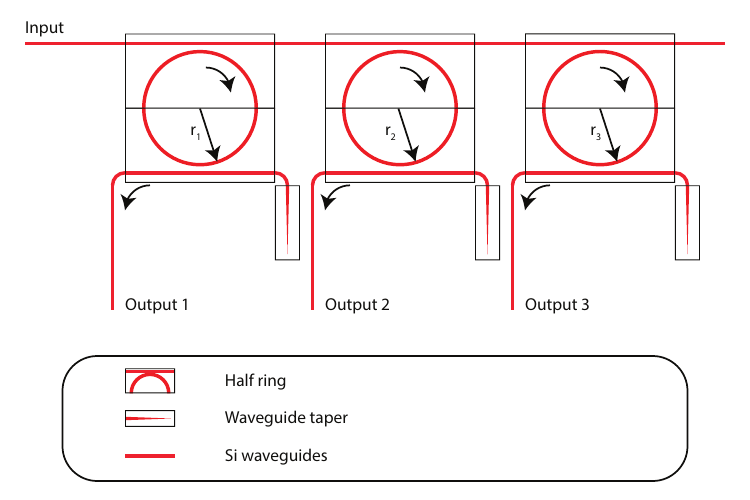}
\caption{A block diagram of a simple add-drop filter made from rings of different radii. The circuit is divided into components that are later connected using sub-network growth. Using our library components built using machine learning models, simulations can be run in quick succession to optimize the radius of each ring to select the desired frequency.}
\label{fig:ring_filter}
\end{figure}

Our last example demonstrates the ability of Simphony to combine resonant devices to create a ring-resonator based add-drop filter, as shown in Figure \ref{fig:ring_filter}.  The circuit consists of three two-port ring-resonators cascade in series, each with a different radius. Each ring has a through port and a drop port. The blocks used in the Simphony calculation are shown in Figure \ref{fig:ring_filter}, and consist of two half-rings coupled to waveguides and a tapered waveguide. Each ring is modeled as its own parameterized subcircuit, having been generated by a function that takes a ring radius as a parameter and returns a subcircuit that can be used within other circuits. The three resulting subcircuits are added to the final circuit to be simulated. A tutorial with code for this example can be found in the online documentation.

The compact model for the waveguide-coupled half-rings were generated using custom machine-learning techniques that will be described in a future manuscript, and model the phase and amplitude response function correctly as verified with FDTD for SiO$_2$-clad silicon rings of arbitrary radius, width, and thickness.  These compact models are included in the Simphony open-source package, and are among the most valuable contributions to the community since ring resonator models are very computationally expensive.  To the authors' knowledge, no other block-driven photonics software includes compact models with the capability to accurately model---within seconds---ring resonators with arbitrary device parameters and coupling geometries.

\begin{figure}[t]
\centering
\includegraphics[width=0.5\textwidth]{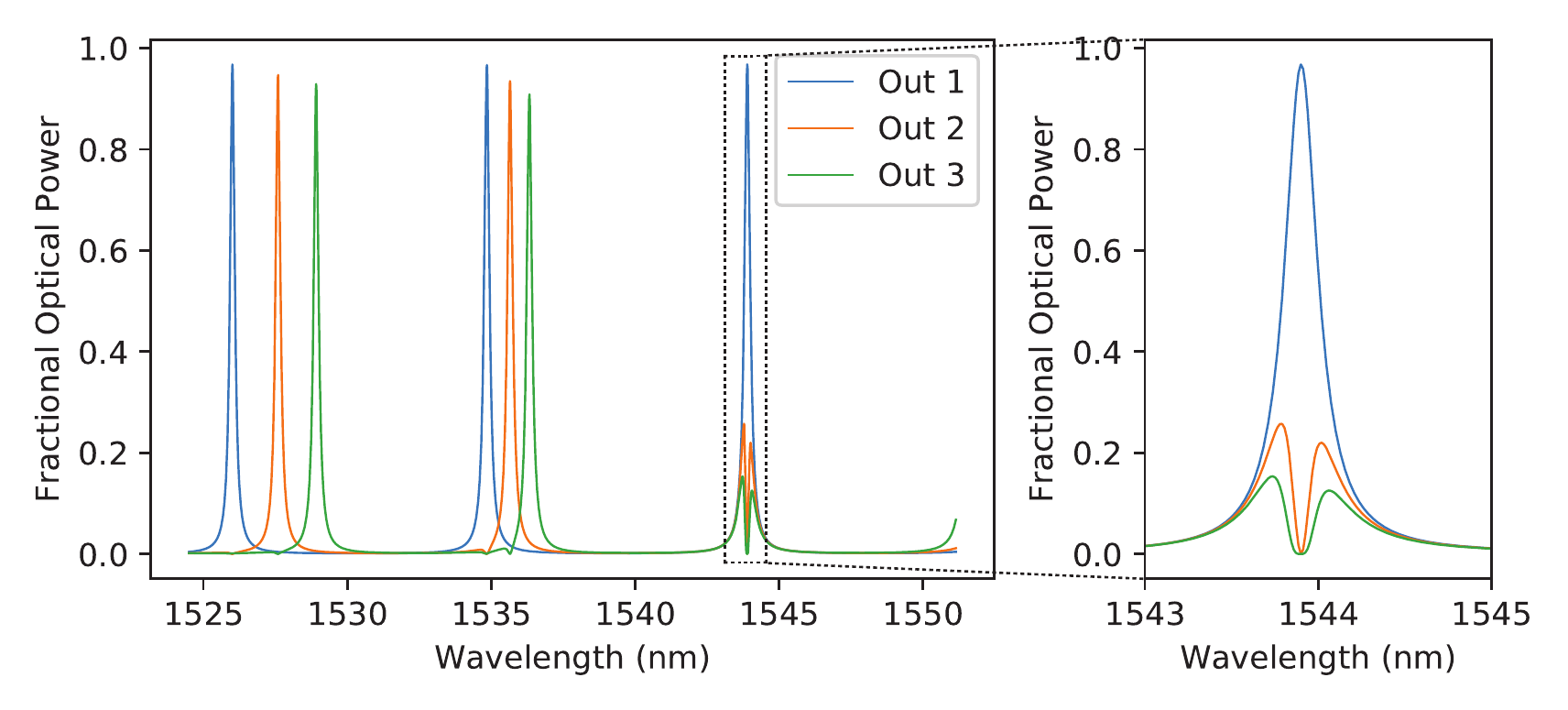}
\caption{Simulation results for the add-drop filter circuit.}
\label{fig:ring_results}
\end{figure}

The simulation results for this example circuit are shown in Figure \ref{fig:ring_results}.  The three rings have radii of $r_1 = 10 \mu$m, $r_2 = 11 \mu$m, and $r_3 = 12 \mu$m.  They thus have different free-spectral ranges and slightly different coupling constants, as can be seen in resulting drop-port power data.  Also, as shown in the resonance near 1545 nm, the causal nature of the optical power flowing from the left to the right is evident where the ring resonances overlap.  The dropped power from the ring with radius $r_1$ maintains a Lorentzian lineshape, while the dropped power from the rings with radius $r_2$ and $r_3$ is depleted near resonance center owing to light already being dropped by the first ring.

\section{Conclusion}
The toolbox presented in this paper is an open-source alternative to PIC design and simulation that allows for the integration of compact models from a variety of different sources. This integration is valuable to designers as a lack of standardization currently exists making interoperability between tools difficult.  The toolbox also happens to be much faster than other currently available tools and allows for easy parallelization and extension.  

We demonstrated the ability of the toolbox to simulate a wide variety of useful integrated photonic circuits, some of which we fabricated and compared to the simulation results.  It is anticipated that this toolbox will be of great utility for researchers seeking a streamlined design interface for photonic integrated circuits as well as for researchers and educators lacking access to expensive commercial software. Moving forward we anticipate collaborative efforts from the open-source community will assist in adding functionality and additional components to the toolbox. 

\section*{Acknowledgment}
The authors would like to acknowledge the University of Washington, the University of British Colombia, and the SiEPIC program for providing resources for the fabrication of devices referenced in this paper.

\bibliographystyle{IEEEtran}
\bibliography{references}

\end{document}